\documentclass[onecolumn]{IEEEtran}
\usepackage[utf8]{inputenc}
\usepackage{float}
\usepackage{graphicx}
\usepackage{hyperref}
\usepackage{xcolor}
\definecolor{lg}{RGB}{220, 220, 220}
\usepackage{listings}
\usepackage{color}
\usepackage{flushend}
\usepackage{multicol}
\usepackage{geometry}

\begin{document}

\title{Implementation of SquashFS Support in U-Boot}
\author{\IEEEauthorblockN{Mariana Villarim \IEEEauthorrefmark{1}, João Marcos Costa \IEEEauthorrefmark{1} and Diomadson Belfort \IEEEauthorrefmark{1} \IEEEmembership{Senior Member - ~IEEE}}
\IEEEauthorblockA{\IEEEauthorrefmark{1} Federal University of Rio Grande do Norte}
\thanks{Corresponding author: Mariana Villarim (email: mariana.villarim1020001@ufrn.edu.br).}}

\maketitle
\begin{abstract}
U-Boot is a notorious bootloader and Open Source project. This work had as objective adding support for the SquashFS filesystem to U-Boot and the support developed was submitted as a contribution to the project. The bootloader is responsible, in this context, for loading the kernel and the device tree blob into RAM. It needs to be capable of reading a storage device's partition formatted with a specific filesystem type. Adding this support allows U-Boot to read from SquashFS partitions. The source code was submitted to U-Boot's mailing list through a series of patches to be reviewed by one of the project's maintainer. Once it gets merged, the support will be used and modified by U-Boot's international community.
\end{abstract}

\section{Introduction}
Embedded systems were called "Stored Program Control" systems in the late 1960s, referring to the memory that held the program and routing information used in communications \cite{williams1999embedding}. When talking about storage in these devices, there are several factors to be analysed, such as capacity, disk speed and, at a higher level, the type of filesystem adopted. A filesystem support is a software that extracts information from a storage device partition formatted with this same filesystem.

SquashFS is distributed as a Linux kernel source patch and is a popular filesystem of the embedded world, and it was not yet
supported by U-Boot. Supporting a filesystem means writing the source code that will bring the necessary logic to parse raw data and provide a regular set of files and directories which can be parsed through a human eye. 

In the embedded Linux context, the platform (e.g. Beagle Bone, Raspberry Pi or any custom board) has its main application running from within a GNU/Linux operating system, as opposed with bare-metal applications. In order to load the Linux kernel, more detailed in the section \ref{sec:kernel}, we expect the bootloader to read a storage medium (e.g. an SD Card) and load the relevant binaries in RAM. 

There are basically two ways to store these files in any storage medium: either by creating partitions bigger than the expected files and writing each file in a separate partition, or creating a partition with a filesystem and storing the binaries as files. The main drawback of the first approach is that no one knows the exact sizes of the binaries, so if a binary grows, it means we must rewrite the entire storage to shift everything.

SquashFS is a compressed and read-only filesystem with performance advantages, such as low overhead, and suited for platforms where memory is constrained. It supports several compression algorithms and has high-speed compression abilities. It is already supported in Linux and the main goal of this work is to add it to the official U-Boot.

Another contribution of this work is a command-line tool, on which one can analyze a SquashFS filesystem image and get various information as the meta-data \footnotemark, the directories and and the content of the files themselves.
\footnotetext{Meta-data is the data which is not the actual payload of the filesystem but instead describes where is the data, what are the names and attributes of each file, etc.}


An embedded Linux application runs under a few conditions: it needs a bootloader to load the kernel and the device tree into RAM, then the kernel needs to mount a filesystem, where the application and its dependencies are stored. The following subsections will detail the terms used to define such sequence.

\subsection{Bare-metal applications}

Bare-metal stands for running applications directly on the hardware level, without the support of an operating system \cite{kirat2014barecloud}. This is commonly achieved by programming a microcontroller directly with C or assembly languages. A bootloader is an example of bare-metal application and most commonly used platforms to develop bare-metal applications are Arduino boards and StMicro's Nucleo \cite{niyonkuru2015bare}. 

\subsection{Embedded Linux}

Embedded Linux is about having an embedded application running on top of an operating system (Linux), unlike bare-metal applications. In this scenario, the platform will embed a distribution (e.g. Debian, Arch) customized to keep only the essential functionalities \cite{lyytinen2009designing}. Common examples of embedded Linux are found in networking devices (e.g. routers and switches), navigation equipment (i.e. GPS), medical instruments, Smart TVs, etc \cite{chen2016towards}. 
Those platforms need more RAM than the ones used to develop bare-metal software. Some of the Linux advantages are that it is highly-tested and stable, then less prone to bugs, its portability (hardware independent), there are free drivers available, which leads to faster development and it is easily scalable.

Bare-metal applications demand fewer system resources and may run much faster, but their complexity level grows exponentially, there is no multi-threading in one core, and some functionalities may need to be adapted to specific hardware \cite{kominos2017bare}. 

\subsection{RAM}

RAM \textit{(Random-access-memory)} is the computer's primary memory resource, used to read and write data. It is said to be random because the data accesses take the same amount of time regardless of the data's physical location, as in opposition to sequential memory access. Since the CPU (Central Processing Unit) can only run programs that have been loaded into RAM (or SRAM - Static Random Acess Memory), in order to load the kernel into RAM, one needs a bootloader.

\subsection{Bootloader}

\begin{figure}[H]
    \centering
    \includegraphics[scale=0.6]{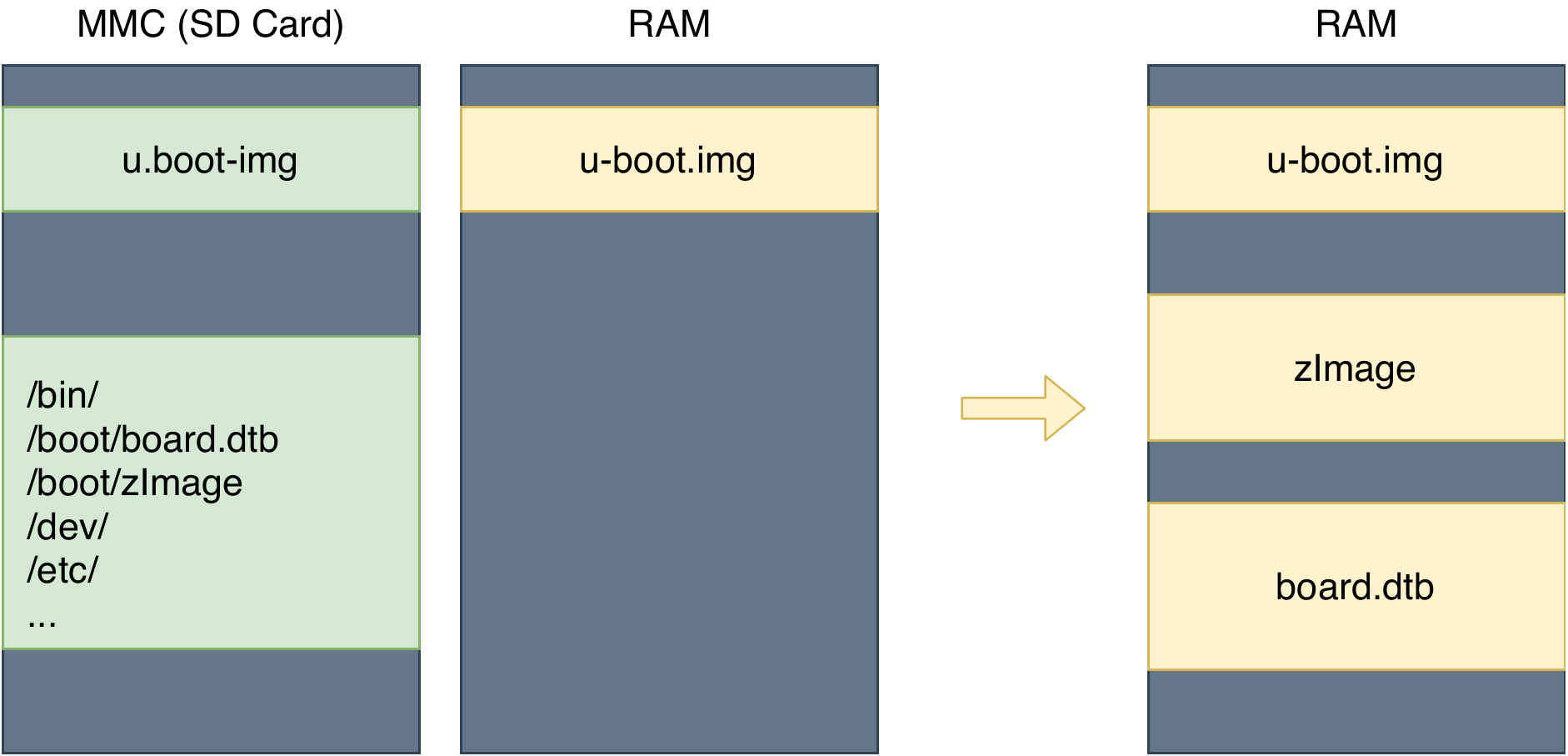}
    \caption{Bootloader scheme}
    \label{fig:bootloader}
\end{figure}

In the embedded Linux context, a bootloader is a program responsible for initializing very few components, loading the necessary binaries to boot the main operating system in RAM and starting it. In our context, the bootloader is called U-Boot and we want to load a kernel (Linux, called zImage) and a device tree (called board.dtb in Figure \ref{fig:bootloader}) describing the hardware peripherals available on the platform. 

The loading operation involves identifying the binaries location, initializing the relevant storage medium, reading the storage medium and extracting the relevant files out of a binary blob when using a filesystem.

\subsection{Kernel} \label{sec:kernel}

\begin{figure}[H]
    \centering
    \includegraphics[scale=0.7]{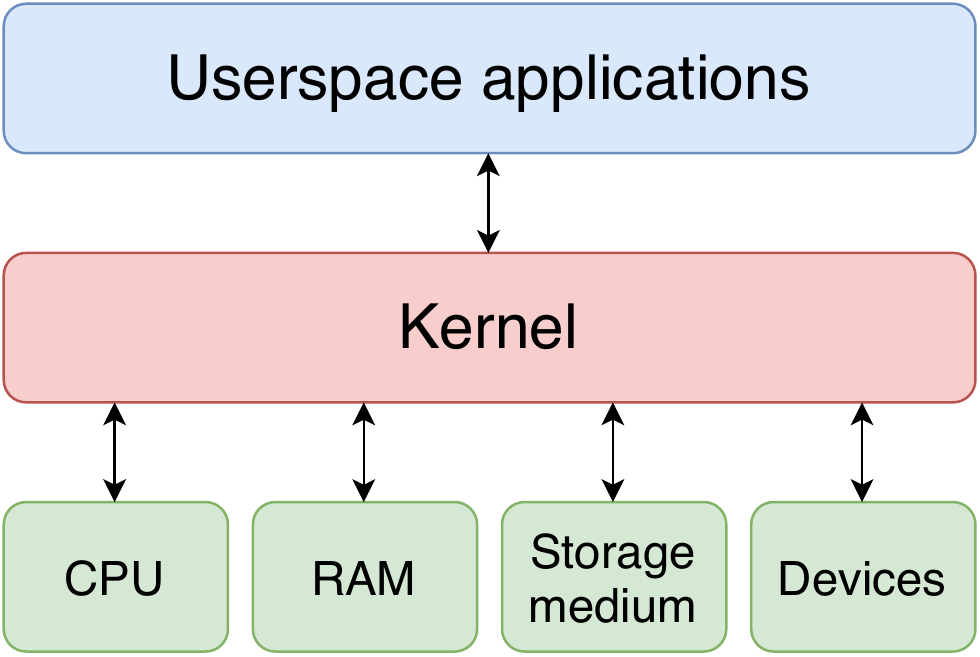}
    \caption{Operating system layers}
    \label{fig:kernel_schematic}
\end{figure}

Kernel is the core program of the operating system \cite{tanenbaum2015modern}. It manages the interactions between software and hardware components like keyboards, monitors, speakers, and memory itself, as shown in Figure \ref{fig:kernel_schematic}. The Linux kernel is modular \cite{love2010linux}, meaning that kernel objects (such as drivers) can be inserted and removed while the kernel is running. The kernel is also responsible for scheduling processes and managing threads.

Kernel mounts the root filesystem after being loaded. This is the stage where the operating system accesses the storage medium (e.g. the SD card), recognizes the filesystem type and, if supported, reads it and makes it available for usage through an interface (i.e. software layer) called VFS (Virtual File System) \cite{kernel}. This VFS allows the filesystem, i.e. the files and directories in the storage medium, to be accessed by userspace programs. 

\subsection{Device tree}

In order to manage the hardware components, the kernel needs a file that describes those components, called Device Tree. It is a tree structure whose nodes represent the physical devices of the system. The bootloader loads the Device Tree Blob (.dtb), the compiled Device Tree. This binary allows the kernel to bind specific devices to their corresponding drivers \cite{corbet2005linux}.

\subsection{Filesystem}

Filesystems are a way to organize data following the structure of a tree. Without a filesystem, the raw data into a storage device cannot be interpreted as files and directories, because there would be no way to tell where each memory range (corresponding to a file) starts and ends. A filesystem stores metadata containing information about its files and directories: their names, sizes, positions, etc. Common filesystem types are: EXT4, NTFS, FAT, Btrfs, UBIFS, SquashFS, etc.

\begin{figure}[H]
    \centering
    \includegraphics[scale=0.7]{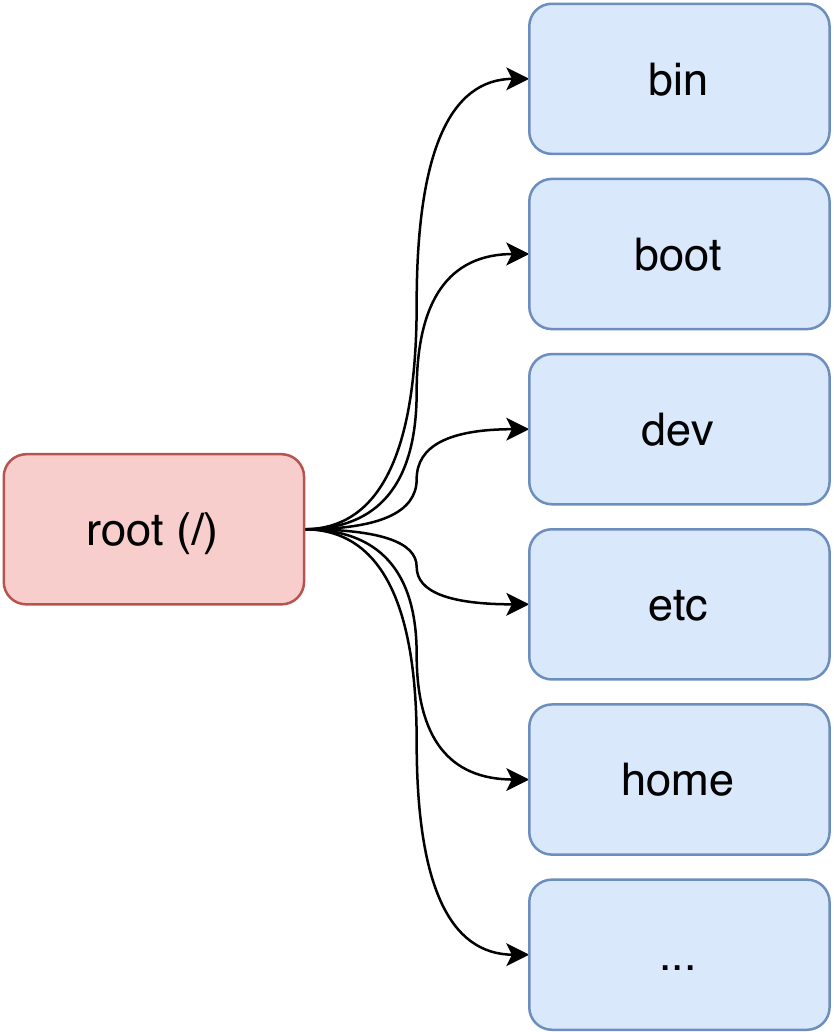}
    \caption{Filesystem Hierarchy Standard}
    \label{fig:FHS}
\end{figure}

In UNIX-like system, such as Linux distributions, the filesystem follows a standard hierarchy of directories (i.e. Filesystem Hierarchy Standard \cite{russell2004filesystem}), starting from a base directory known as root and branching out into other directories with standardized names and purposes, as shown in Figure \ref{fig:FHS}

\subsection{SquashFS} \label{sec:sqfs}

SquashFS is a widely spread, compressed, and read-only filesystem that offers performance attributes \cite{squashfs1}. The data and metadata can be stored compressed, using compression algorithms such as GZIP, LZO, ZSTD, etc. SquashFS can be used for general purposes, both in desktop computers and embedded platforms. It has high-speed decompression abilities, and it is specially useful in contexts where data must not be overwritten, such as in Secure Boot. SquashFS is also intended for situations where storage resources are limited.

\begin{figure}[H]
    \centering
    \includegraphics[width=\linewidth]{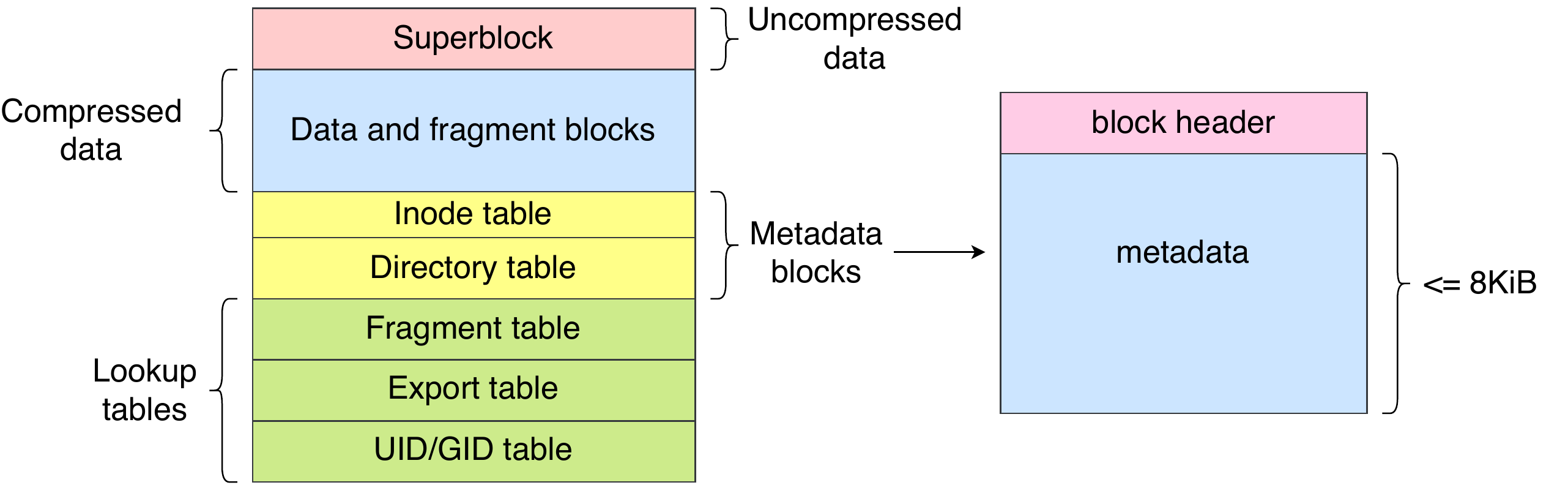}
    \caption{SquashFS image layout}
    \label{fig:sqfs}
\end{figure}

Linux mainline kernel supports SquashFS since 2009 and Barebox, another bootloader for embedded platforms, supports SquashFS as well. SquashFS image is divided in specific sections, as the Figure \ref{fig:sqfs} illustrates. Those sections are usually compressed, except for the superblock. Each section contains different types of information.

\begin{enumerate}
    \item Superblock: it is the first section and indicates the compression type used, where the other sections begin and other information about the archive.
    \item Data and fragment blocks: the data from the files that compose the SquashFS image. These data are stored sequentially in pre-fixed size blocks. If the file cannot be fully stored in blocks (i.e. its size is not equally divisible by the block size), the file tail may be stored in a fragment block. 
    \item Inode table: every file and directory is represented by an inode holding information as the file size, the symbolic link target, the inode type, etc. The inodes are stored into a sequence of metadata blocks.
    \item Directory table: information about the directories is also present in the directory table, where each directory is represented by one or more headers followed by a list of its contents. The directories and files names are stored into the directory table.
    \item Fragment table: this table contains the blocks and offsets where the files fragments are stored. It does not contain the fragments themselves.
    \item Export table: it contains inode references for faster lookups in cases as NFS (Network File System). This table is optional, and can be set when compiling the SquashFS image.
    \item UID/GID table: stores files users and group identifiers.
\end{enumerate}

\subsection{U-Boot}

U-Boot is probably the most popular bootloader for embedded platforms, and it is an open source project as well. It supports filesystems types such as FAT, EXT, and UBIFS \cite{gediya2019open}. 
Adding the support for SquashFS is the main point of this research. U-Boot has a command-line interface where the user can run specific commands, both in the host machine and in the target board. Concerning filesystems, U-Boot's two most important commands are \textit{ls}, to display a directory's content, and \textit{load}, to load a file in memory, at an specific address. Each filesystem names them differently, prefixing the command with the filesystem name (e.g. \textit{fatload} and \textit{fatls} for FAT). The SquashFS support offers the user these very same commands, so it can navigate into an SquashFS formatted partition, list the files inside and load them.

\subsection{Contributing to open source projects}

As previously mentioned, Linux and U-Boot are examples of open source projects. This means that the source code of the project is available for execution, inspection, modification, and redistribution. The source code is stored in the cloud, publicly available, and developers can contribute by sending patches to the mailing list of the project. Patches are files generated to describe what has changed between two versions of a project: what has been deleted, added or modified. These patches will be reviewed by the project's community, who will either accept the patch series or send it back with comments to be addressed.

\section{Materials and Methods}
\section{SquashFS Support Development}

This section details the tasks executed in order to accomplish the last version of the SquashFS support.
The development stage can be divided into two parts: firstly a user-space command-line tool, then the actual development over U-Boot.

\subsection{User-space Command-line Tool}

The development started with a simple tool called \textit{squashfs-utils}, detailed in the following subsections, to understand the SquashFS organization.

\subsubsection{Command-line Parsing}

The project started by writing a Makefile, a simple \textit{README}, and then adding a license file (i.e. GPL v2.0). 
The goal was to separate the command-line parsing from the actual SquashFS functionalities, because U-Boot has its own parsing logic.

In the main function, a logic to analyze the command-line was written, retrieving what option had been used and if there were any parameters.

\subsubsection{SquashFS Sections Dumping}

Three dumping options were implemented, corresponding to the superblock, the inode table, and the directory table. The output format for each of these options corresponds to the layout of each section on-disk, which means that the inodes, for example, are displayed in the same order they are stored on-disk.

Dumping the superblock was the easier task, because this section always has the same size and only one structure is needed to retrieve its information. However, the directory and inode tables were more complex.

Those tables can be compressed, so zlib was used for decompression. Also, the inodes, directory headers and entries do not have a fixed length, so the loop used to iterate through these tables needed to calculate each item's size and then increment a pointer with it.

These functions have only debugging and comprehension purposes, and one could use them to inspect SquashFS compiled images. Critical parts were extracted from their scopes in order to be used in the actual U-Boot support, but the code sections responsible for actually printing information were discarded later on.

The \textit{squashfs-utils} usage message is shown in Listing \ref{usage}:
\lstset{caption={squashfs-utils usage message},label=usage}
\begin{lstlisting}
$ ./sqfs -h
usage: sqfs [-h]
       sqfs [-s] [-i] [-d] <fs-image>
       sqfs [-e] <fs-image> /path/to/dir/
       sqfs [-e] <fs-image> /path/to/file

Tool to analyze the content of a SquashFS image

Options:
       -h: Prints the usage and exits
       -s: Dumps the contents of a SquashFS image's
       superblock
       -i: Dumps the contents of a SquashFS image's
       inode table
       -d: Dumps the contents of a SquashFS image's
       directory table
       -e: Dumps the contents of a SquashFS image's
       file or directory.
	   For directories, end path with '/'.

Parameters:
       <fs-image>: Path to the filesystem image
\end{lstlisting}

The following examples dump a SquashFS image generated from \textit{source-dir} directory. It contains another directory (empty), a text file and a symbolic link to the file, as we can see in Listing \ref{source-dir}. The image is compiled using \textit{mksquashfs}.

\lstset{caption={Source directory content},label=source-dir}
\begin{lstlisting}
$ tree source-dir 
source-dir
+-- dir_example
+-- file.txt
+-- slink -> file.txt

1 directory, 2 files
\end{lstlisting}

The output of \textit{squashfs-utils} when the superblock option is used is shown in Listing \ref{superblock}:
\lstset{caption={Superblock information},label=superblock}
\begin{lstlisting}
$./sqfs -s source-dir.sqfs 
--- SUPER BLOCK INFORMATION ---
Magic number: sqsh
Number of inodes: 4
Filesystem creation date: Tue 2020-08-04
(yyyy-mm-dd) 15:46:14 CEST
Block size: 131 kB
Number of fragments: 1
Block log: 17
Compression type: ZLIB
Super Block Flags: 0xc0
Major/Minor numbers: 4/0
Root inode: 0x60
Bytes used: 312
Id table start: 0x130
(xattr) Id table start: 0xffffffffffffffff
Inode table start: 0x6c
Directory table start: 0xbc
Fragment table start: 0x105
Lookup table start: 0x122
--- SUPER BLOCK FLAGS ---
Duplicates
Exportable
\end{lstlisting}

The magic number will be used later, into the U-Boot support, to check if the device's partition actually contains a SquashFS image. The number of inodes is coherent: 2 directories (\textit{dir\_example} and the root directory), the symbolic link and the regular file. Since the text file is too small to be stored into a whole 131 KB data block, a fragment is used instead. This can be verified by looking at the number of fragments and the bytes used by the image.

The inode table dump result is detailed in Listing \ref{inodetable}:
\lstset{caption={Inode table information},label=inodetable}
\begin{lstlisting}
$./sqfs -i source-dir.sqfs
--- --- ---
{Inode 1/4}
--- --- ---
Permissions: 0x01fd
UID index: 0x0000
GID index: 0x0000
Modified time: Tue 2020-08-04
(yyyy-mm-dd) 15:41:41 CEST
Inode number: 1
Inode type: Basic Directory
Start block: 0x00000000
Hard links: 2
File size: 3
Block offset: 0x0000
Parent inode number: 4

{Inode 2/4}
--- --- ---
Permissions: 0x01b4
UID index: 0x0000
GID index: 0x0000
Modified time: Tue 2020-08-04
(yyyy-mm-dd) 15:42:17 CEST
Inode number: 2
Inode type: Basic File
Start block: 0x00000000
Fragment block index: 0x00000000
Fragment block offset: 0x00000000
(Uncompressed) File size: 12

{Inode 3/4}
--- --- ---
Permissions: 0x01ff
UID index: 0x0000
GID index: 0x0000
Modified time: Tue 2020-08-04
(yyyy-mm-dd) 15:45:48 CEST
Inode number: 3
Inode type: Basic Symlink
Hard links: 1
Symlink size: 8
Target path: file.txt

{Inode 4/4}
--- --- ---
Permissions: 0x01fd
UID index: 0x0000
GID index: 0x0000
Modified time: Tue 2020-08-04
(yyyy-mm-dd) 15:45:48 CEST
Inode number: 4
Inode type: Basic Directory
Start block: 0x00000000
Hard links: 3
File size: 63
Block offset: 0x0000
Parent inode number: 5

\end{lstlisting}

Inodes are stored with their numbers in an ascending order, and the last inode corresponds to the root directory. Every directory inode (even the root) has a parent, and the inode 1 parent is the root (inode 4). 

Next, there is the text file's inode. Its content is "Hello world", making a total of 11 characters plus the terminator character, wich leads to a file size of 12 bytes. The symbolic link is also coherent, since both \textit{tree} command and the inode dump show the same target path. 

The Listing \ref{dirtable} contains the directory table dump:

\lstset{caption={Directory table information},label=dirtable}
\begin{lstlisting}
$./sqfs -d source-dir.sqfs
Directory 1
Name: dir_example
Empty directory.

Root directory
1) dir_example
2) file.txt
3) slink

\end{lstlisting}

This option displays the directories in the order they are placed in the SquashFS image. Below the directory number, its name and content are displayed. The content is stored in alphabetical order.

Finally, there is also the option to dump a single file's content, as shown in Listing \ref{singlefile}:
\lstset{caption={Dumping file's content},label=singlefile}
\begin{lstlisting}
$ cat source-dir/file.txt 
Hello world

$./sqfs -e source-dir.sqfs /file.txt 
Hello world
\end{lstlisting}



\subsection{Development over U-Boot}

Before migrating the previous code, this section introduces U-Boot's organisation and build system, aiming to know where to place files, what Makefiles to edit and what build options to declare.

\subsubsection{U-Boot's build system and Sandbox architecture}

\begin{figure}
    \centering
    \includegraphics[width=\linewidth]{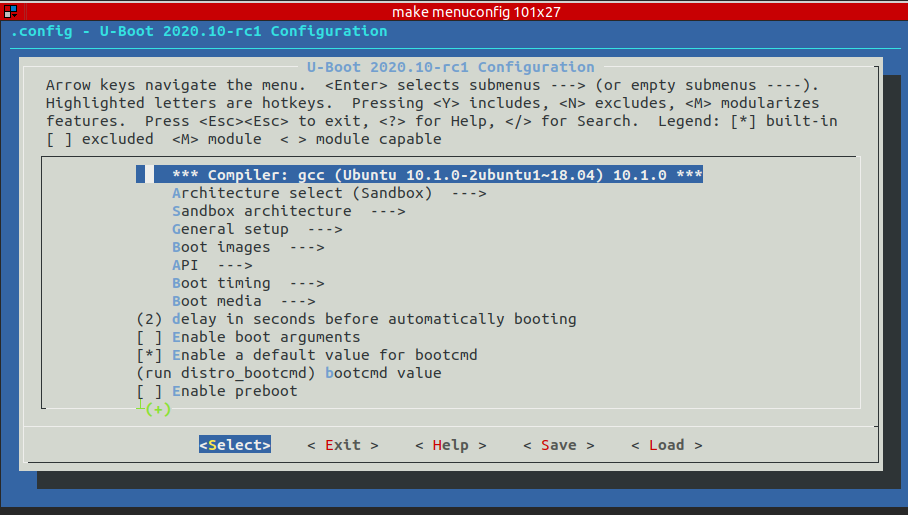}
    \caption{Menuconfig interface.}
    \label{fig:menuconfig}
\end{figure}

U-Boot uses KBuild \cite{kbuild}, the same build (and configuration) system as the Linux kernel. It provides an infrastructure to select what modules will be compiled and linked to the final binary, and it depends on the standard GNU's make. The configurations (or \textit{configs}) can be selected with the \textit{menuconfig} interface, as show in Figure \ref{fig:menuconfig}.

The built U-Boot binary can be tested in userspace, without the need to cross-compile to a target board. This is achieved by compiling 'Sandbox' architecture, as shown in Listing \ref{sandbox}, which is hardware independent and can run under Linux, working as a normal C application. Running the built result, we have U-Boot's console:

\lstset{caption={U-Boot's sandbox console},label=sandbox}
\begin{lstlisting}
U-Boot 2020.10-rc1  Testing SquashFS support
- by Joao Marcos Costa  
-00154-gc7b2d6a45d (Aug 04 2020 - 18:10:16 +0200)

DRAM:  128 MiB
WDT:   Not found!
MMC:   
In:    serial
Out:   serial
Err:   serial
SCSI:  
Net:   No ethernet found.
Hit any key to stop autoboot:  0 
=> 
\end{lstlisting}

\subsubsection{Migrating the code from squashfs-utils}

\begin{figure}
    \centering
    \includegraphics[scale=0.6]{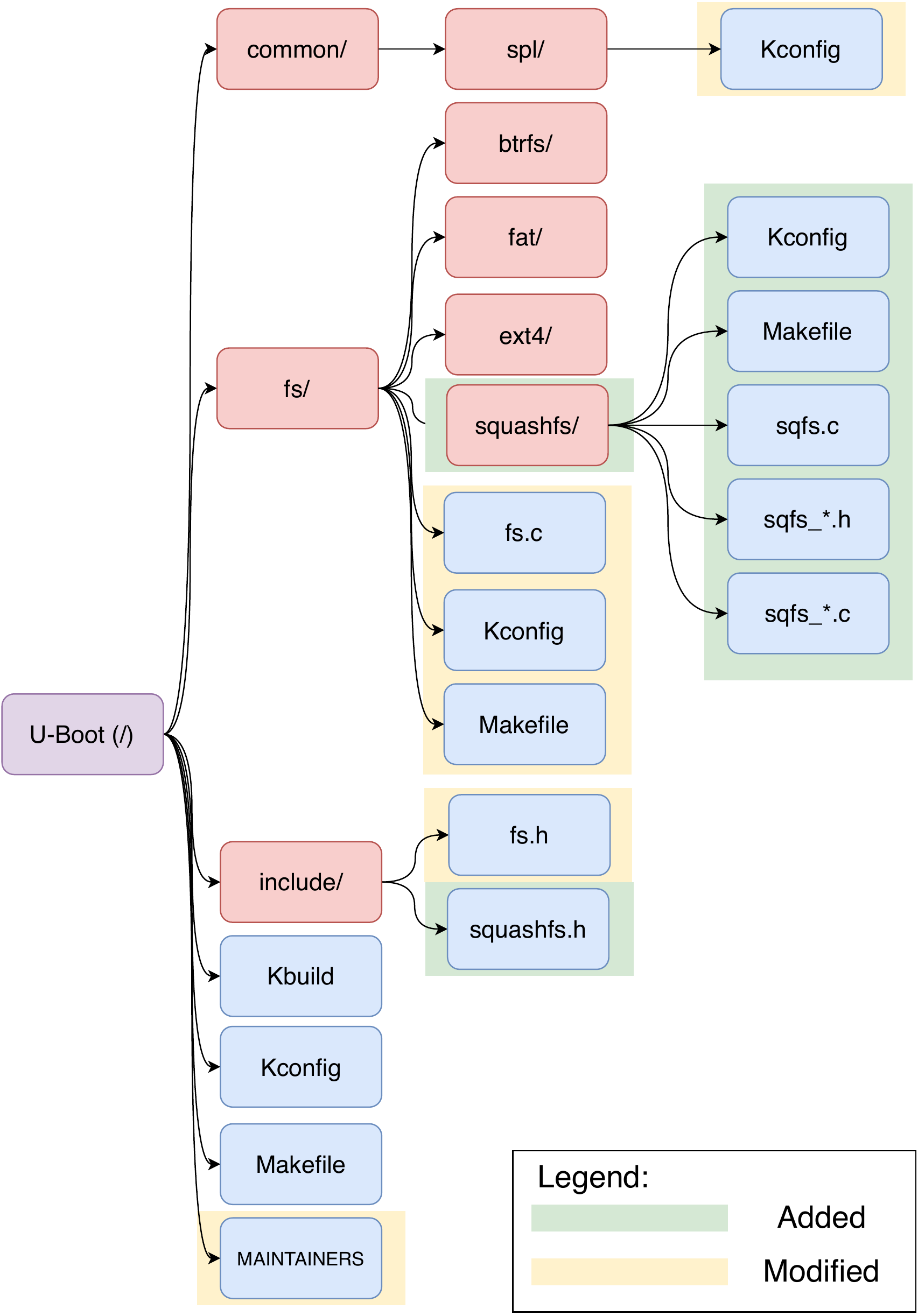}
    \caption{Adding the new filesystem to U-Boot.}
    \label{fig:patch1}
\end{figure}

The migration started by creating empty headers and source files in the same directory as the other filesystems are placed, as shown in the Figure \ref{fig:patch1}. U-Boot's tree is extensive, so the figure represents pertinent files and directories only. The new files and directories are highlighted in green, and the modified ones are highlighted in yellow.

Adapting the code previously written consisted in assuring the following criteria:
\begin{enumerate}
    \item Replacing userspace memory mapping by direct memory access to read memory blocks from the storage device containing the SquashFS image.
    \item Removing the command-line parsing code and the dumping functions.
    \item Prefixing structures, macros, and functions with "sqfs\_".
    \item Removing any unused code section.
\end{enumerate}

The \textit{MAINTAINERS} file in the project's root directory contains a list of contributors responsible for specific sections of the project. It is recommended to add your name and the new section (files and directories) once you add new files to the project.

\subsubsection{Implementing U-Boot's API functions}

U-Boot uses a polymorphic strategy to support various filesystems, which is based in function pointers to which there are filesystem-specific implementations assigned. This API goes beyond the following list, but these were the functions needed for SquashFS support:

\begin{enumerate}
    \item \textit{probe}: it is the first function to be called when a filesystem command is executed. In the SquashFS implementation, it checks if the partition contains a SquashFS image and fills a global context structure describing the disk partition, the current device and the image's superblock. 
    \item \textit{ls}: lists the files in a given directory.
    \item \textit{size}: gets the size of a file. It is always called before \textit{read}.
    \item \textit{read}: reads the file content into a buffer and tells how many bytes were read.
    \item \textit{close}: clears the global context structure filled by \textit{probe}. It needs to be implemented because it is always called at the end of a filesystem command, but it does not have a standard utility and some filesystems leave it with an empty definition.
    \item \textit{opendir}: opens a directory stream and returns a pointer to it. 
    \item \textit{readdir}: iterates through every entry in the directory stream.
    \item \textit{closedir}: frees memory allocated in \textit{opendir}.
\end{enumerate}

Those functions are listed here with their original API's names, but they were implemented prefixed by "sqfs\_".

\subsubsection{Implementing the SquashFS commands}

At first, it was implemented \textit{ls} for SquashFS. It consisted simply in calling the API functions implemented, so the function's definition was short and straightforward. However, if the API implementation previously done was actually correct, we would be able to use the generic version of \textit{ls} provided by U-Boot, without the need to implement \textit{sqfs\_ls}.

Therefore, this observation made us investigate and find troubles in the implementation and in some structures we had defined. After fixing all these issues, we removed \textit{sqfs\_ls} and started using the generic \textit{ls}.

The \textit{read} function is, basically, a continuation of \textit{size}. It checks if the file is stored in both data and fragment blocks, or only in one of these. Then, it decompresses the blocks into a buffer and counts how many bytes were read.

The commands need an usage message and a config to enable them in Kbuild. Later on, scripts were added to test those commands. After enabling the commands with the \textit{menuconfig} interface, they can be executed in the U-Boot's console as in Listing \ref{cmd}:
\lstset{caption={SquashFS commands},label=cmd}
\begin{lstlisting}
U-Boot 2020.10-rc1  Testing SquashFS support
- by Joao Marcos Costa
-00154-gc7b2d6a45d (Aug 04 2020 - 18:10:16 +0200)

DRAM:  128 MiB
WDT:   Not found!
MMC:   
In:    serial
Out:   serial
Err:   serial
SCSI:  
Net:   No ethernet found.
Hit any key to stop autoboot:  0 
=> sqfsls 
sqfsls - List files in directory. Default: root (/).

Usage:
sqfsls <interface> [<dev[:part]>] [directory]
    - list files from 'dev' on 'interface' in
    'directory'

=> sqfsload 
sqfsload - load binary file from a SquashFS
filesystem

Usage:
sqfsload <interface> [<dev[:part]> [<addr>
[<filename> [bytes [pos]]]]]
    - Load binary file 'filename' from 'dev' on 
      'interface' to address 'addr' from SquashFS
      filesystem. 'pos' gives the file position to
      start loading from. If 'pos' is omitted, 0
      is used. 'pos' requires 'bytes'. 'bytes'
      gives the size to load. If 'bytes' is 0 or
      omitted, the load stops on end of file.
      If either 'pos' or 'bytes' are not aligned
      to ARCH_DMA_MINALIGN then a misaligned
      buffer warning will be printed and
      performance will suffer for the load.
\end{lstlisting}

\begin{figure}[H]
    \centering
    \includegraphics[scale=0.6]{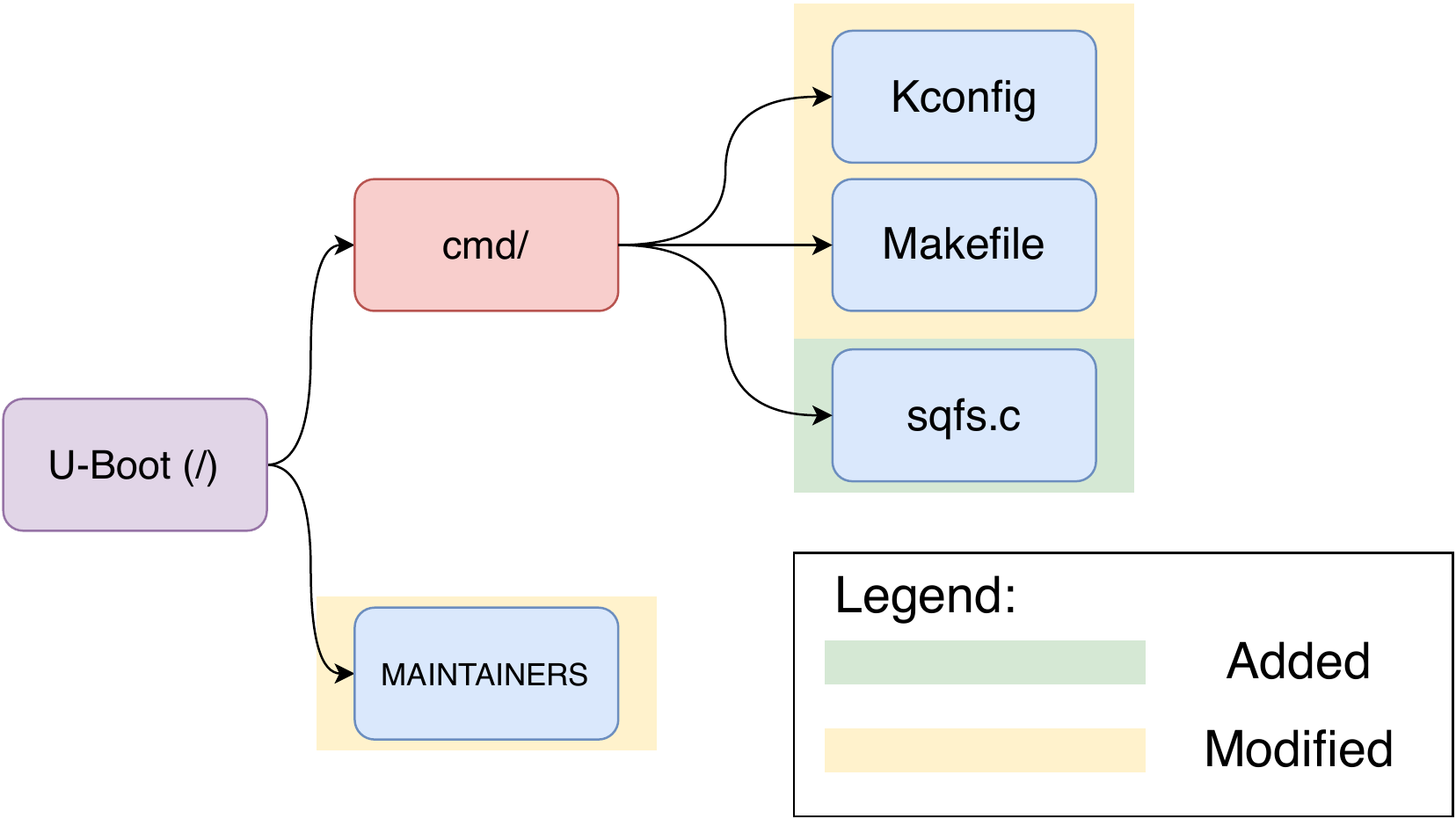}
    \caption{Adding commands to U-Boot.}
    \label{fig:commands}
\end{figure}

\section{Results and Discussion}
Three important challenges were faced while developing this filesystem driver were: enabling the use of symbolic links, segmentation faults, and data aborts.

The segmentation faults were due to memory leaks and an erroneous memory management. We started compiling Sandbox with GDB flags and then running it under Valgrind. This helped us to find every memory leak and every call to \textit{malloc()} that did not have its corresponding \textit{free()}.

SquashFS supports symbolic links, and we were advised that they were necessary and it should be implemented. This task consisted in a lot of string parsing and recursion, and string parsing helpers are usually prone to bugs. Finally, the symbolic links worked fine and without any memory leak.

The data aborts were the most hard to find because Valgrind does not identify them, since they appear only in the target board, not in userspace. This error is reported by U-Boot in cases as segmentation faults and unaligned memory accesses. All the segmentation fault occurrences were solved, but we did not know the concept of unaligned memory access. It is needed to review the whole code to find places where this could happen and correct the memory accesses using the proper Linux solutions provided by its source code (which is available in U-Boot's sources), such as the macro \textit{get\_unaligned()}.

\subsection{Adding test scripts}

It is recommended to add a new test every time a new command is added. U-Boot's tests are written in C, Shell and Python, and the long term goal is to have all the tests written in Python due to its simplicity and flexibility. In this work, all the scripts are in Python. 

The tests run U-Boot (Sandbox), execute given commands and retrieve the console's output. Then, multiple assertions are made to check if a certain string is present in the command's output.

These tests are responsible for generating a directory with a few files of random content inside. Then, it calls \textit{mksquashfs} and generates an SquashFS image. The idea is to have three files with a fixed size so the SquashFS image has one file stored only in data blocks, a second one stored only in a fragment block, and a third file stored in both data and fragment blocks. Finally, there is also a symbolic link. These files are loaded and the number of bytes read is asserted with the file size (previously known). When the test is over or when an exception is raised, the generated files are deleted.

\begin{figure}
    \centering
    \includegraphics[scale=0.5]{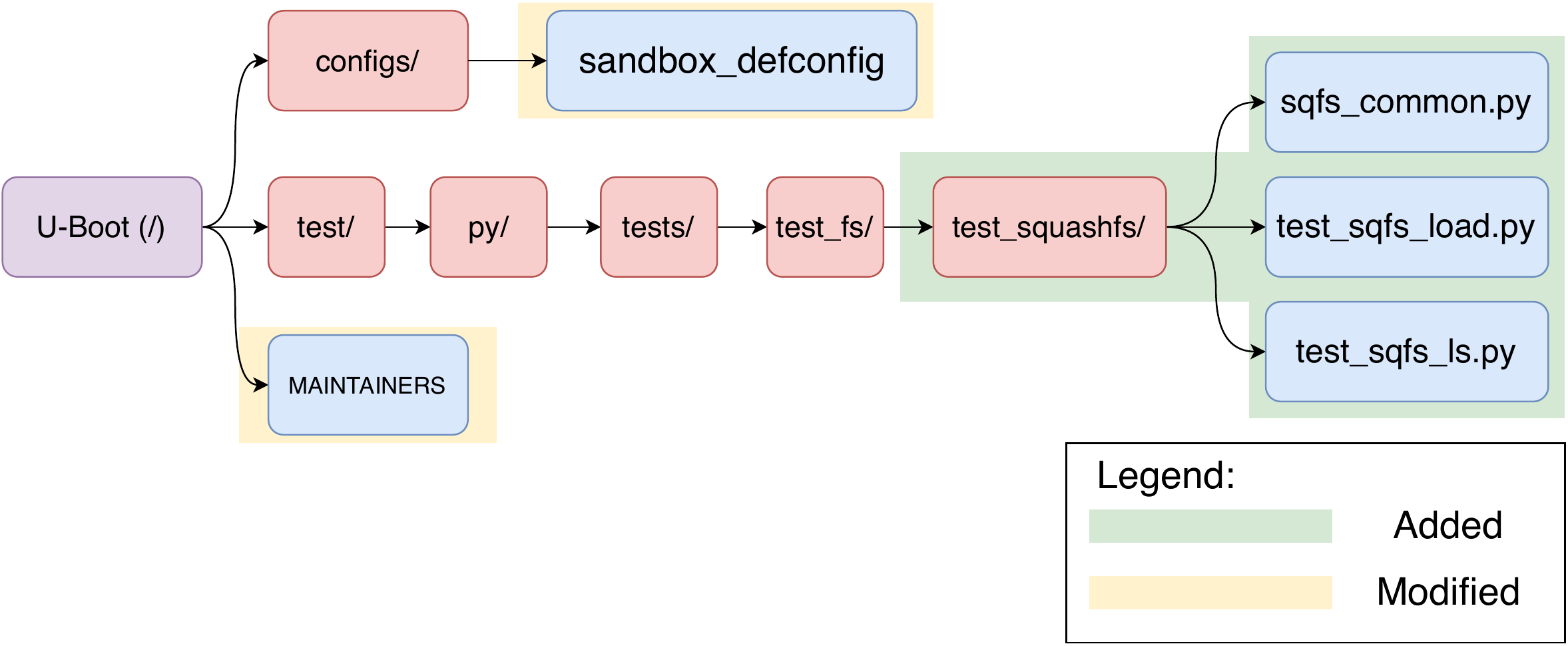}
    \caption{Adding test scripts to U-Boot.}
    \label{fig:tests}
\end{figure}

In U-Boot, the test files and directories must be prefixed with "test\_". U-Boot has its own API to send commands and retrieve their outputs. It also allows to check if a config is enabled and if the host machine has a specific dependency. In this case, the dependency was \textit{mksquashfs}, a tool to generate SquashFS images. The configs needed were \textit{FS\_SQUASHFS} and \textit{CMD\_SQUASHFS}: the filesystem support and its commands, respectively.

We were recommended to add \textit{CMD\_SQUASHFS} to Sandbox's configuration file. This means that Sandbox will have the SquashFS commands by default and they will not need to be enabled manually. Since those were new files to the project, we needed to update the \textit{MAINTAINERS} file as well. 

\subsection{Submitting the support}
As it was previously mentioned, contributions to open source projects as U-Boot are made through a mailing list, where patches are sent and reviewed. 

Extensive contributions should be divided into a series of patches, where each patch concerns an specific subject, so the contribution will be easier to review. Also, the series should start by a cover letter presenting the series's purpose and what files will be touched by the patches. The series in this paper consisted of:

\begin{enumerate}
    \item \textit{Add support for the SquashFS filesystem}: the cover letter.
    \item \textit{fs/squashfs: new filesystem}: adds the sources for SquashFS support. It is the largest patch, containing the SquashFS structures, API's implementation, etc.
    \item \textit{fs/squashfs: add filesystem commands}: adds the commands \textit{ls} and \textit{load}.
    \item \textit{include/u-boot, lib/zlib: add sources for zlib decompression}. This patch adds an extract of Zlib's source code so we could use a particular function that allows decompressing a buffer into another.
    \item \textit{fs/squashfs: add support for zlib decompression}. Adds a call to zlib's previously added function. This modification was made in a different section of U-Boot's code, so it was separated from the previous patch.
    \item \textit{fs/fs.c: add symbolic link case to fs\_ls\_generic()}. Since we used the generic \textit{ls} provided by U-Boot, we needed to make a small change to this function's definition. It is only used by FAT filesystem, where there is not a concept of symbolic link. This is a trivial patch, adding simply three lines to the function, in order to display symbolic links aside from regular files and directories.
\end{enumerate}

The patch title starts by what directories it touches, followed by a sentence in the imperative form summarising what the patch does. The patch also contains the commit message, giving more details on how it was conceived and what changes it brings. If successive versions were sent, the patch should also contain a "change log" that accumulates what has been changed since the first version until the current one.

Patches should not contain style issues. There is a Perl script called \textit{checkpatch} in U-Boot's sources that parses text files and displays style issues of three kinds: errors, warnings, and checks. Also, the code added by the patch series cannot give warnings of any kind nor errors when compiled.


The SquashFS support allows any board to be booted with a kernel and device tree present in a SquashFS image. First, the Listing \ref{sqfsls} shows the U-Boot console and the result of the \textit{sqfsls} command:

\lstset{caption={sqfsls command},label=sqfsls}
\begin{lstlisting}
U-Boot SPL 2020.10-rc1-00154-gc7b2d6a45d
(Aug 07 2020 - 11:17:02 +0200)
Trying to boot from MMC1

U-Boot 2020.10-rc1-00154-gc7b2d6a45d
(Aug 07 2020 - 11:17:02 +0200)

CPU  : AM335X-GP rev 2.1
Model: TI AM335x BeagleBone Black
DRAM:  512 MiB
WDT:   Started with servicing (60s timeout)
NAND:  0 MiB
MMC:   OMAP SD/MMC: 0, OMAP SD/MMC: 1
Loading Environment from FAT...
Unable to use mmc 0:1... <ethaddr> not set.
Validating first E-fuse MAC
Net:   Could not get PHY for
ethernet@4a100000: addr 0
eth2: ethernet@4a100000, eth3: usb_ether
Hit any key to stop autoboot:  0 
=> sqfsls mmc 0:1
            bin/
            boot/
            dev/
            etc/
            lib/
    <SYM>   lib32
    <SYM>   linuxrc
            media/
            mnt/
            opt/
            proc/
            root/
            run/
            sbin/
            sys/
            tmp/
            usr/
            var/

2 file(s), 16 dir(s)
\end{lstlisting}

Then, the commands in Listing \ref{boot} are executed on a Beagle Bone Black Wireless (e.g. arm architecture) to start the kernel:
\lstset{caption={Starting the kernel in a Beagle Bone Black},label=boot}
\begin{lstlisting}
=> sqfsload mmc 0:1 $kernel_addr_r /boot/zImage
6091376 bytes read in 476 ms (12.2 MiB/s)
=> sqfsload mmc 0:1 0x81000000
/boot/am335x-boneblack.dtb
40817 bytes read in 14 ms (2.8 MiB/s)
=> setenv bootargs console=ttyO0,115200n8
=> bootz $kernel_addr_r - 0x81000000
## Flattened Device Tree blob at 81000000
   Booting using the fdt blob at 0x81000000
   Loading Device Tree to 8fff3000, end 8fffff70 ...
   OK

Starting kernel ...

[    0.000000] Booting Linux on physical CPU 0x0
[    0.000000] Linux version 4.19.79
(joaomcosta@joaomcosta-Latitude-E7470)
(gcc version 7.3.1 20180425 [linaro-7.3-2018.05
revision d29120a424ecfbc167ef90065c0eeb7f91977701]
(Linaro GCC 7.3-2018.05))
#1 SMP Fri May 29 18:26:39 CEST 2020
[    0.000000] CPU: ARMv7 Processor [413fc082]
revision 2 (ARMv7), cr=10c5387d

\end{lstlisting}

Those are just the first messages sent by the kernel, but they are an evidence that the SquashFS image was successfully decompressed and read. 

\section{Conclusions}

Developing this filesystem driver needed a special attention to, firstly, the filesystem on-disk layout, then to U-Boot's architecture. It showed how the project's constraints change when it is cross-compiled and when tests are run with bigger SquashFS images.

The personal ideas of coding style and software engineering must be set aside when contributing to an Open Source project, because the project already has its own coding style and architecture to be respected.

Reviewing major contributions such as this filesystem support takes longer than small patches. The first version was sent on July 7th, 2020 and the contribution was finally merged on August 8th, 2020 at its fourth version. Once the support is merged, U-Boot's community is able to use and improve it. New compression algorithms can be added, the string parsing can be simplified, the search for inodes and directories can be optimized, etc.

The project developed in this paper can be accessed through Github link \url{https://github.com/u-boot/u-boot}. This directory contains the source code for U-Boot, a bootloader for Embedded boards based on PowerPC, ARM, MIPS and several other processors, which can be installed in a boot ROM and used to initialize and test the hardware or to download and run application
code.

\bibliographystyle{IEEEtran}
\bibliography{arxiv}

\end{document}